\documentclass[12pt]{amsart}
\usepackage{amssymb}
\usepackage{color}
\usepackage{rotating}
\newtheorem{claim}{}[section]
\newtheorem{theorem}[claim]{Theorem}

\newtheorem{lemma}[claim]{Lemma}

\begin{document}
\baselineskip 6.2 truemm
\parindent 1.5 true pc

\newcommand\lan{\langle}
\newcommand\ran{\rangle}
\newcommand\tr{{\text{\rm Tr}}\,}
\newcommand\ot{\otimes}
\newcommand\wt{\widetilde}
\newcommand\join{\vee}
\newcommand\meet{\wedge}
\renewcommand\ker{{\text{\rm Ker}}\,}
\newcommand\im{{\text{\rm Im}}\,}
\newcommand\mc{\mathcal}
\newcommand\transpose{{\text{\rm t}}}
\newcommand\FP{{\mathcal F}({\mathcal P}_n)}
\newcommand\ol{\overline}
\newcommand\JF{{\mathcal J}_{\mathcal F}}
\newcommand\FPtwo{{\mathcal F}({\mathcal P}_2)}
\newcommand\hada{\circledcirc}
\newcommand\id{{\text{\rm id}}}
\newcommand\tp{{\text{\rm tp}}}
\newcommand\pr{\prime}
\newcommand\e{\epsilon}
\newcommand\inte{{\text{\rm int}}\,}
\newcommand\ttt{{\text{\rm t}}}
\newcommand\spa{{\text{\rm span}}\,}
\newcommand\conv{{\text{\rm conv}}\,}
\newcommand\rank{\ {\text{\rm rank of}}\ }
\newcommand\vvv{\mathbb V_{m\meet n}\cap\mathbb V^{m\meet n}}
\newcommand\ppp{\mathbb P_{m\meet n} + \mathbb P^{m\meet n}}
\newcommand\re{{\text{\rm Re}}\,}
\newcommand\la{\lambda}
\newcommand\msp{\hskip 2pt}
\newcommand\ppt{\mathbb T}
\newcommand\rk{{\text{\rm rank}}\,}
\def\cc{\mathbb{C} }
\def\pp{\mathbb{P} }
\def\rr{\mathbb{R} }
\def\qq{\mathbb{Q} }
\def\a{\alpha }
\def\b{\beta }

\title{Exposedness of Choi type  entanglement witnesses and applications to lengths of separable states}

\author{Kil-Chan Ha}
\address{Faculty of Mathematics and Statistics, Sejong University, Seoul 143-747, Korea}

\author{Seung-Hyeok Kye}
\address{Department of Mathematics and Institute of Mathematics\\Seoul National University\\Seoul 151-742, Korea}

\thanks{KCH is partially supported by NRFK 2012-0002600. SHK is partially supported by NRFK 2012-0000939}



\begin{abstract}
We present a large class of indecomposable exposed positive linear maps
between three dimensional matrix algebras.
We also construct two qutrit separable states with lengths ten in the
interior of their dual faces. With these examples, we show that the length of a separable state
may decrease strictly when we mix it with another separable state.
\end{abstract}

\maketitle

\section{Introduction}

Entanglement is the main resource in the current quantum information
theory as well as quantum physics, and the distinction of
entanglement from separability is one of the main research topics.
The most complete method to detect entanglement is to use the
duality theory \cite{eom-kye,horo-1} between separable states and
positive linear maps in matrix algebras, and this was formulated as
the notion of entanglement witnesses \cite{terhal}. By the duality,
an entanglement witness is just a positive linear map which is not
completely positive, under the Jamio\l kowski-Choi isomorphism
\cite{choi75-10,jami}. The duality naturally gives rise to the
notion of exposedness. A positive map $\phi$ is said to be exposed
if its bidual consists of the nonnegative scalar multiplications of
itself. An exposed map automatically generates an extreme ray of
the convex cone $\mathbb P$ consisting of
all positive linear maps.

Main tasks to distinguish entanglement from separability is to detect entanglement with positive partial transposes
due to the PPT criterion \cite{choi-ppt,peres} for separability, and
we need indecomposable positive maps to detect those entanglement.
Entanglement witnesses arising from exposed indecomposable positive
maps play very important role in this situation.
First of all, it was observed in \cite{ha+kye_exposed} that
the set of all exposed indecomposable entanglement witnesses is
enough to detect every entangled PPT state.
Note that the dual face of an exposed map must be maximal among
nontrivial faces of the convex cone $\mathbb S$ of all unnormalized separable states.
Furthermore, if an
indecomposable entanglement witness $W$ is exposed then
the set of PPT entangled states detected by $W$ has an nonempty interior. Note that every extremal
decomposable map is already exposed by \cite{marcin_exp}.

Importance of exposed positive maps also arises from the notion of optimality
of entanglement witnesses. An entanglement witness $W$ is said to be optimal \cite{lew00}
if the set of entanglement detected by $W$ is maximal with respect to the set inclusion.
It should be noted that an indecomposable optimal entanglement witness $W$ need not detect a
maximal set of PPT entangled states \cite{ha-kye-optimal}. In other to detect a maximal set of
PPT entangled states, the partial transpose $W^\Gamma$ should be optimal as well as $W$.
We call $W$ co-optimal if $W^\Gamma$ is optimal. In the language of positive maps, it turns out that
a positive map $\phi$ is optimal (respectively co-optimal) if and only if the smallest face
of the cone $\mathbb P$ containing $\phi$ has no completely positive (respectively copositive) map.
See expository article \cite{kye-ritsu}.
More stronger condition than optimality is given by the spanning property which is easier to check, and it turns out that
every indecomposable exposed positive maps enjoy both spanning property and co-spanning property.
Therefore, the notion of exposedness gives rise to the very strong optimality of entanglement witnesses.
It should be noted that the mere extremeness need not imply the spanning property, as the Choi map
\cite{choi-lam} shows.

After the paper \cite{ha+kye_exposed}, many efforts \cite{chru,chru--W,chru--R,sar-chru,sengupta}
have been made to understand the structures of exposed indecomposable positive linear maps. See also \cite{majew,majew-ty}.
Nevertheless, there are still very few known examples of exposed indecomposable positive maps.
Actually, exposed positive maps given in \cite{ha+kye_exposed} are only known examples
for $3\otimes 3$ cases in the literature, to the best knowledge of the authors.
These maps have been used in \cite{sko-3x3} to show that there is no \lq easy\rq\ criterion for separability in the $3\otimes 3$ cases.

In this paper, we consider the entanglement witnesses in \cite{ha+kye_Choi}, to show that some of them are exposed.
This is given by
\[
W[a,b,c;\theta]=\left(
\begin{array}{ccccccccccc}
a     &\cdot   &\cdot  &\cdot  &-e^{i\theta}     &\cdot   &\cdot   &\cdot  &-e^{-i\theta}     \\
\cdot   &c &\cdot    &\cdot    &\cdot   &\cdot &\cdot &\cdot     &\cdot   \\
\cdot  &\cdot    &b &\cdot &\cdot  &\cdot    &\cdot    &\cdot &\cdot  \\
\cdot  &\cdot    &\cdot &b &\cdot  &\cdot    &\cdot    &\cdot &\cdot  \\
-e^{-i\theta}     &\cdot   &\cdot  &\cdot  &a     &\cdot   &\cdot   &\cdot  &-e^{i\theta}     \\
\cdot   &\cdot &\cdot    &\cdot    &\cdot   &c &\cdot &\cdot    &\cdot   \\
\cdot   &\cdot &\cdot    &\cdot    &\cdot   &\cdot &c &\cdot    &\cdot   \\
\cdot  &\cdot    &\cdot &\cdot &\cdot  &\cdot    &\cdot    &b &\cdot  \\
-e^{i\theta}     &\cdot   &\cdot  &\cdot  &-e^{-i\theta}     &\cdot   &\cdot   &\cdot  &a
\end{array}
\right)
\]
with nonnegative $a,b,c$ and real $\theta$,
which is the Choi matrix of the linear map defined by
\[
\Phi[a,b,c;\theta](X)=
\left(
\begin{array}{ccc}
ax_{11}+bx_{22}+cx_{33} & -e^{i\theta}x_{12} & -e^{-i\theta}x_{13} \\
-e^{-i\theta}x_{21} & cx_{11}+ax_{22}+bx_{33} & -e^{i\theta}x_{23} \\
-e^{i\theta}x_{31} & -e^{-i\theta}x_{32} & bx_{11}+cx_{22}+ax_{33}
\end{array}
\right),
\]
for $X=[x_{ij}]\in M_3$. When $\theta=0$, these give rise to just positive maps
considered in \cite{cho-kye-lee}. We refer to \cite{ha+kye_Choi} for
various subcases with specific $a,b,c$ and $\theta$ considered in the
literature. These maps have been constructed to find entanglement witnesses detecting $3\otimes 3$
PPT entangled edge states \cite{kye_osaka} of type $(8,6)$, whose existence was a long-standing question.

It was shown in \cite{ha+kye_Choi} that the map $\Phi[a,b,c;\theta]$ is positive if and only if two conditions
$a+b+c\ge p_{\theta}$ and $bc\ge (1-a)^2$ are satisfied,
where
\[
p_{\theta}=\max\left\{2\cos \left(\theta+\frac{2\pi}3\right),\, 2\cos \theta,\,
2\cos \left (\theta-\frac{2\pi} 3\right)\right\}.
\]
The interesting cases occur when
\[
2-p_{\theta}\le a< 1,\qquad a+b+c= p_{\theta},\qquad bc= (1-a)^2.
\]
These cases are parameterized by
\[
a(t)=1-\frac{(p_{\theta}-1)t}{1-t+t^2},\quad b(t)=\frac{(p_{\theta}-1)t^2}{1-t+t^2},\quad
c(t)=\frac {(p_{\theta}-1)}{1-t+t^2}
\]
with $0<t<\infty$, and we write
\[
\Phi_{\theta}(t)=\Phi[a(t),b(t),c(t);\theta].
\]
In \cite{ha+kye_exposed}, it was shown that $\Phi_{0}(t)$ is exposed
whenever $t\neq 1$. In this paper, we show that $\Phi_{\theta}(t)$
is also exposed whenever the condition
\begin{equation}\label{cond-exp}
\theta \neq\frac {2n-1}3\pi,\qquad  (\theta, t)\neq \left(\frac {2n}3\pi,\, 1\right)
\end{equation}
holds, under which it is known \cite{ha+kye_Choi} that
$\Phi_{\theta}(t)$ is indecomposable, and has the bi-spanning property.

We note that some of $\Phi[a,b,c;\theta]$ give rise to concrete counterexamples
to the structural physical approximation conjecture raised in \cite{korbicz}.
See \cite{ha_kye_spa, stormer_spa}.
Especially, it was shown in \cite{ha_kye_spa} that some of $\Phi_{\theta}(t)$ violate the SPA conjecture.
Therefore, we conclude that even exposed positive map may violate this conjecture.
Note that exposed decomposable positive maps satisfy the SPA conjecture by \cite{aug_bae}.

We also construct a separable state $\varrho_0$ in the interior of the dual face of
$\Phi_\theta(t)$, whose length is $10$, which exceeds $3\times 3$.
The length $\ell(\varrho)$ is defined by the smallest number $k$ with
which $\varrho$ is the convex combination of $k$ pure product
states \cite{DiVin,p-horo}. So, the length of a separable state represents the minimal
physical efforts that realize the state.
It was asked in \cite{chen_dj_semialg}
if there exist a separable state $\sigma$ and a state $\rho=\sigma+|\xi\otimes \eta\rangle\langle \xi \otimes \eta|$ such that
$\ell(\rho)<\ell(\sigma)$. We construct a separable state $\varrho_1$ such that $\ell(\varrho_0+\varrho_1)=9$
to give a partial negative answer to this question.

In the next section, we review very briefly duality between the cone $\mathbb P$ and the cone $\mathbb S$, together with
the duality between the convex cone $\mathbb D$ of all decomposable maps
and the convex cone $\mathbb T$ generated by all bipartite states
with positive partial transposes. We also explain the relations between
exposedness and several optimality conditions including the spanning property.
After we show the exposedness of the map $\Phi_{\theta}(t)$ in Section 3,
the length of a separable state in the dual face will be calculated in Section 4.
Throughout this paper, $|\bar x\rangle$ denotes the vector whose entries are the complex conjugates of the corresponding entries
of the vector $|x\rangle$ and $V^*$ denotes the Hermitian transpose of the matrix $V$.

\section{Duality}

For a linear map $\phi:M_m\to M_n$ from the matrix algebra $M_m$ of all $m\times m$ matrices into $M_n$ and a bipartite
state $\varrho\in M_m\otimes M_n$, the bilinear pairing $\langle\varrho,\phi\rangle$ is defined by
\[
\langle\varrho,\phi\rangle=\text{\rm tr}(\varrho C_\phi^{\rm t}),
\]
where $C_\phi^{\rm t}$ is the transpose of the Choi matrix $C_\phi=\sum_{i,j=1}^m e_{ij}\otimes\phi(e_{ij})\in M_m\otimes M_n$
of the map $\phi$, with the usual matrix units $e_{ij}=|i\rangle \langle j|$.
For a pure product state
$\varrho=|\xi\otimes\eta\rangle \langle\xi\otimes\eta|\in M_m\otimes M_n$, the above pairing
may be expressed by
\begin{equation}\label{eq:Choi_pair}
\langle\varrho,\phi\rangle =\langle \bar \eta|\phi(|\xi\rangle \langle \xi|)|\bar\eta\rangle
=\langle \xi\otimes \eta |C^{\rm t}_{\phi}|\xi \otimes \eta\rangle
=\langle \bar \xi\otimes \bar \eta |C_{\phi}|\bar \xi\otimes \bar \eta\rangle.
\end{equation}
The cones $\mathbb P$ and $\mathbb S$ are dual to each other in the following sense:
\[
\begin{split}
\varrho\in\mathbb S \ &\Longleftrightarrow\ \langle\varrho,\phi\rangle \ge 0 \quad {\rm for \ every}\ \phi\in\mathbb P,\\
\phi\in\mathbb P \ &\Longleftrightarrow\ \langle\varrho,\phi\rangle \ge 0 \quad  {\rm for \ every}\ \varrho\in\mathbb S.
\end{split}
\]
Every $\phi\in\mathbb P$ determines the dual face $\phi^\prime$ of the cone $\mathbb S$ given by
\[
\phi^\prime=\{\varrho\in\mathbb S: \langle\varrho,\phi\rangle =0\}.
\]
Every $\varrho\in\mathbb S$ also gives rise to the dual face $\varrho^\prime$ of the cone $\mathbb P$ in the same way.
A face $F$ of a convex cone is said to be exposed if it is a dual face. This is equivalent to say that the bidual
of an interior point of $F$ coincides with $F$.
It is well known that not every face of $\mathbb P$ is exposed, but it is not known yet if every face of $\mathbb S$
is exposed or not. We say that a positive map $\phi$ is exposed if the ray generated by $\phi$ is an exposed face.

By the relation (\ref{eq:Choi_pair}), we note that the dual face $\phi^\prime$ of a positive map $\phi$ is determined by the set
\[
P_\phi:=\{|\xi\otimes \eta\rangle : \phi(|\xi\rangle \langle\xi|)|\bar\eta\rangle=0\}
\]
of product vectors. Therefore,
the exposedness of a positive linear map $\Phi$ is equivalent to the following: If $\phi$
is a positive map satisfying the property
\begin{equation}\label{duble-con}
\langle |\xi\otimes \eta\rangle \langle \xi\otimes \eta|,\phi\rangle=0 \quad {\rm for\ every}\ |\xi\otimes \eta\rangle \in P_\Phi,
\end{equation}
then $\phi$ is a nonnegative scalar multiplication of $\Phi$.

Recall that a positive map $\phi$ has the spanning property \cite{lew00} if the set $P_\phi$ of product vectors spans the whole space
$\mathbb C^m\otimes\mathbb C^n$.
The spanning property is stronger than optimality, and easier to check.
It is said to have the co-spanning property \cite{ha-kye-optimal}
if the composition $\phi\circ {\rm t}$ with the transpose map $\rm t$ has the spanning property. In terms of entanglement witnesses,
an entanglement witness
has the co-spanning property if and only if its partial transpose has the spanning property.
It turns out that $\phi$ has the spanning (respectively co-spanning)
property if and only if the smallest exposed face
of the cone $\mathbb P$ containing $\phi$ has no completely positive (respectively copositive) maps.
See Section 8 of \cite{kye-ritsu}. Therefore, we have the following implications for indecomposable positive maps:

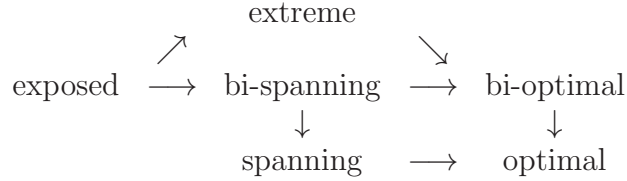
\begin{table}[h!]
\begin{tabular}{ccccc}
                       &                              & {\rm extreme}         &                          &                           \\
   & $\nearrow$   &  & $\searrow$   &   \\
{\rm exposed}  &$\longrightarrow$& {\rm bi-spanning} &$\longrightarrow$& {\rm bi-optimal}\\
                        &  &$\downarrow$&  &$\downarrow$\\
                        &  & {\rm spanning} &$\longrightarrow$& {\rm optimal}\\
\end{tabular}
\end{table}

Recall that we say that a positive map has the bi-spanning  property
if it has both spanning property and co-spanning property, and is
bi-optimal if it is both optimal and co-optimal. We have examples
\cite{ha-kye-optimal,ha+kye_Choi} which show that no implication can be reversed
in the above rectangle parts of implications concerning various notions on optimality and the spanning property.
See also \cite{ha-yu} for an example of a bi-optimal witness
which is not extreme. Note that the original Choi map \cite{choi-lam} has not the spanning property even if it is extreme.  Finally, 
we note that the positive map $\Phi_0 (1/b)+\Phi_0(1/b)\circ {\rm t}$ is not extreme, but it has the bi-spanning property for each $b\neq 1$ as it was implicitly observed in Section 5 of \cite{ha+kye_unique_decom}. 

Now, we turn our attention to the duality between the convex cone $\mathbb D$ of all decomposable maps and the cone
$\mathbb T$ generated by all PPT states. They are dual to each other, in the same way as for the cones $\mathbb P$ and $\mathbb S$.
We consider the elementary operators given by
\[
\phi_V:X\to V^* X V,\qquad \phi^W:X\to W^* X^{\rm t}W
\]
for $m\times n$ matrices $V$ and $W$, which are typical examples of completely positive maps and completely copositive maps.
For finite families ${\mathcal V}=\{V_i\}$ and ${\mathcal W}=\{W_i\}$ of $m\times n$ matrices, we put
\[
\phi_{\mathcal V}+\phi^{\mathcal W}=\sum \phi_{V_i}+\sum \phi^{W_j}.
\]
A positive map is said to be decomposable if it is of the form $\phi_{\mathcal V}+\phi^{\mathcal W}$.
For a pure product state $\varrho$ given by
$|\xi\otimes\eta\rangle \langle\xi\otimes\eta|$, we have
\begin{equation}\label{eq:pair}
\begin{aligned}
\langle \varrho,\phi_V\rangle
=&\langle \bar \eta|V^*|\xi\rangle \langle \xi|V|\bar \eta\rangle
=|\langle \xi|V|\bar \eta\rangle|^2,\\
\langle \varrho,\phi^W\rangle
=&\langle \bar \eta|W^* (|\xi\rangle \langle \xi|)^{\rm t}W|\bar \eta\rangle
=\langle \bar \eta|W^*|\bar \xi\rangle \langle \bar \xi| W|\bar \eta\rangle
=|\langle \bar \xi |W|\bar \eta\rangle|^2.
\end{aligned}
\end{equation}

\section{Exposedness}

In this section, we show that the positive map $\Phi_\theta(t)$ is exposed. First of all, we examine the positive map $\Phi_\theta(t)$ under the following condition
\begin{equation}\label{cond}
-\frac\pi 3<\theta< \frac\pi 3,\qquad  (\theta,t)\neq (0,1).
\end{equation}
We first recall \cite{ha+kye_Choi} that
all product vectors $|\xi \otimes \eta\rangle$ in $P_{\Phi_{\theta}(t)}$ are of the form
\begin{equation}\label{eq:case1}
|z_1(a_1,a_2,a_3)\rangle:=\begin{cases}
(a_1,a_2,a_3)^{\rm t} \otimes (\bar a_1,\bar a_2e^{-\frac{2\pi}3i},\bar a_3e^{\frac{2\pi}3i})^{\rm t}&(-\pi<\theta<-\frac{\pi}3)\\
(a_1,a_2,a_3)^{\rm t} \otimes (\bar a_1,\bar a_2,\bar a_3)^{\rm t}&  (-\frac{\pi}3<\theta<\frac{\pi}3)\\
(a_1,a_2,a_3)^{\rm t} \otimes (\bar a_1,\bar a_2e^{\frac{2\pi}3i},\bar a_3e^{-\frac{2\pi}3i})^{\rm t}&  (\frac{\pi}3<\theta<\pi)
\end{cases}
\end{equation}
with $ |a_1|=|a_2|=|a_3|$,
\begin{eqnarray}
\label{eq:case2}
|z_2(\beta)\rangle&:=& (0,\sqrt{t}\beta,1)^{\rm t}\otimes (0,\sqrt{t} \bar \beta, t e^{i\theta})^{\rm t}\quad {\rm with }\ |\beta|=1,\\
\label{eq:case3}
|z_3(\beta)\rangle&:=&(1,0,\sqrt{t} \beta)^{\rm t}\otimes (t e^{i\theta},0,  \sqrt{ t}\bar\beta)^{\rm t} \quad {\rm with }\ |\beta|=1,\\
\label{eq:case4}
|z_4(\beta)\rangle&:=&(\sqrt{t}\beta,1,0)^{\rm t}\otimes (\sqrt t\bar \beta,t e^{i\theta}, 0)^{\rm t} \quad {\rm with }\ |\beta|=1.
\end{eqnarray}

In order to exploit product vectors of the form (\ref{eq:case2}), (\ref{eq:case3}) and (\ref{eq:case4}),
we first consider product states $\varrho$ given by
\begin{equation}\label{eq:casexx}
(\sqrt t\beta,1)^{\rm t}\otimes (\sqrt t \bar\beta,te^{i\theta})^{\rm t}\quad {\rm with }\ |\beta|=1.
\end{equation}
in $\mathbb C^2\otimes\mathbb C^2$, and determine positive linear maps $\phi: M_2\to M_2$ such that $\langle\varrho,\phi\rangle=0$
for every $\varrho$ given by the product vectors of the form (\ref{eq:casexx}).
Recall that $\phi$ is decomposable by \cite{woronowicz}, and so it must be of the form $\phi=\sum\phi_{V_i}+\sum\phi^{W_j}$
for $2\times 2$ matrices $V_i$ and $W_j$, for which we have $\langle\varrho,\phi_{V_i}\rangle=\langle\varrho,\phi^{W_j}\rangle=0$.
By (\ref{eq:pair}),
we see that $\phi$ is of the form
$\phi=\alpha \phi_V+\alpha^\prime \phi^W$ for nonnegative $\alpha$ and $\alpha^\prime$, with the matrices
$V=\left(\begin{array}{cc}1&0\\0&-e^{i\theta}\end{array}\right)$ and
$W=\left(\begin{array}{cc}0&e^{i \theta}\\-t&0\end{array}\right)$.
The Choi matrix of this map is given by
\[
\left(\begin{array}{cccc}
\alpha  &\cdot&\cdot&-\alpha e^{i\theta} -\alpha^\prime  t e^{i\theta} \\
\cdot&\alpha^\prime &\cdot&\cdot\\
\cdot&\cdot& t^2\alpha^\prime &\cdot\\
-\alpha e^{-i\theta}-\alpha^\prime t e^{-i\theta} &\cdot&\cdot&\alpha  \end{array} \right),
\]
which is uniquely decomposed by
$$
\left(\begin{array}{cccc}
\alpha  &\cdot&\cdot&-\alpha e^{i\theta}\\
\cdot&\cdot &\cdot&\cdot\\
\cdot&\cdot&\cdot &\cdot\\
-\alpha e^{-i\theta} &\cdot&\cdot&\alpha  \end{array} \right)
+
\left(\begin{array}{cccc}
\cdot   &\cdot&\cdot& -\alpha^\prime t e^{i\theta} \\
\cdot&\alpha^\prime &\cdot&\cdot\\
\cdot&\cdot& t^2\alpha^\prime &\cdot\\
-\alpha^\prime t e^{-i\theta} &\cdot&\cdot&\cdot   \end{array} \right)
$$
as the sum of a positive matrix and a co-positive matrix,
where $\cdot$ denotes zero. We recall that
a matrix in $M_m\otimes M_n$ is co-positive if it is the partial transpose $X^{\rm t}\otimes Y$ of a positive matrix $X\otimes Y$.

Now, we consider the $9\times 9$ matrix  which is the  Choi matrix
\[
C_\phi=\left(
\begin{array}{ccc}
\phi_{11} & \phi_{12} & \phi_{13}\\
\phi_{21} & \phi_{22} & \phi_{23}\\
\phi_{31} & \phi_{32} & \phi_{33}
\end{array}
\right)
\]
of a positive linear map $\phi$
satisfying the condition (\ref{duble-con})
with $\Phi=\Phi_\theta(t)$.
We note that product vectors of the form (\ref{eq:case4}) determine the $4\times 4$ principal submatrix
with the $1,2,4,5$ columns and rows. Product vectors of the forms (\ref{eq:case2}) and (\ref{eq:case3})
also determine $4\times 4$ principal submatrices, and so we see that $C_\phi$ is of the form
\begin{equation}\label{form-1}
\left(
\begin{array}{ccccccccc}
\alpha  &\cdot&\cdot    &    \cdot & -(\alpha+tp) e^{i\theta}& *    &    \cdot &* & -(\alpha+tr) e^{-i\theta}\\
\cdot&p &*    &    \cdot & \cdot & *    &    * & * &  *\\
\cdot&*&t^2r    &    * & * & *    &    \cdot &* & \cdot\\
\cdot&\cdot&*    &  t^2p & \cdot & *  &   *&* & *\\
-(\alpha +tp) e^{-i\theta}&\cdot&*    &    \cdot & \alpha  & \cdot    &    * &\cdot & -(\alpha +tq) e^{i\theta}\\
*& *& *    &    * & \cdot & q    &  * &\cdot & \cdot\\
\cdot& *&\cdot    &     * & * &  *    &    r &* & \cdot\\
*& *&*    &    * & \cdot & \cdot    &   * &t^2q & \cdot\\
-(\alpha +tr) e^{i\theta}& *&\cdot    &   * & -(\alpha +tq) e^{-i\theta}& \cdot    &    \cdot &\cdot & \alpha  \end{array}
\right),
\end{equation}
where $*$ is to be determined. To do this, we need the following simple lemma:

\begin{lemma}\label{lemma}
Suppose that the matrices
\[
\left(\begin{array}{ccc}
\alpha & \cdot & -\alpha e^{i\theta}\\
\cdot &x &y\\
-\alpha e^{-i\theta}& \bar y&\alpha
\end{array}\right),\qquad
\left(\begin{array}{ccc}
p & v&-tp\, e^{-i\theta}\\
\bar v& u &w\\
-tp \, e^{i\theta}&\bar w&t^2p
\end{array}\right)
\]
are positive. Then we have $y=0$ and $w=-\bar v t e^{-i\theta}$.
\end{lemma}

We consider the $2\times 2$ principal sub-block of (\ref{form-1})
from the first and second blocks, which represents a positive linear
map from $M_2$ into $M_3$. Since this is decomposable by
\cite{woronowicz}, it is the sum of a positive matrix and a
co-positive matrix. We write this decomposition by
\[
\left(\begin{array}{cc} \phi_{11} & \phi_{12}\\ \phi_{21} & \phi_{22}\end{array}\right)
=\left(\begin{array}{cc} P_{11} & P_{12} \\ P_{21} & P_{22} \end{array}\right)
+\left(\begin{array}{cc} Q_{11} & Q_{12} \\ Q_{21} & Q_{22} \end{array}\right).
\]
By the discussion above,
we see that the decomposition of $4\times 4$ principal submatrix from $1,2,4,5$-entries is uniquely determined:
\[
\begin{split}
&\left(
\begin{array}{cccccc}
\alpha  &\cdot&\cdot    &    \cdot & -(\alpha +tp) e^{i\theta}& *  \\
\cdot&p &*    &    \cdot & \cdot & *    \\
\cdot&*&t^2r    &    * & * & *    \\
\cdot&\cdot&*    &  t^2p & \cdot & *  \\
-(\alpha +tp) e^{-i\theta}&\cdot&*    &    \cdot & \alpha  & \cdot     \\
*& *& *    &    * & \cdot & q
\end{array}
\right)\\
=&
\left(
\begin{array}{cccccc}
\alpha  &\cdot&*    &    \cdot & -\alpha e^{i\theta}& *    \\
\cdot&\cdot &*    &    \cdot & \cdot & *    \\
*&*&*    &   * & * & *    \\
\cdot&\cdot&*    &  \cdot & \cdot & *    \\
-\alpha e^{-i\theta}&\cdot&*   &    \cdot & \alpha  & *     \\
*& *& *    &    *& * & *
\end{array}
\right)
+
\left(
\begin{array}{cccccc}
\cdot &\cdot&*    &    \cdot & -tp\, e^{i\theta}& *    \\
\cdot&p &*    &    \cdot & \cdot & *    \\
*&*&*    &   * & * & *    \\
\cdot&\cdot&*    &  t^2p & \cdot & *    \\
-tp\, e^{-i\theta}&\cdot&*   &    \cdot & \cdot  & *     \\
*& *& *    &    *& * & *
\end{array}
\right)
\end{split}
\]
Now, we determine the unknowns as follows:
\begin{itemize}
\item
Use the positivity and co-positivity,
\item
Compare the coefficients on both sides,
\item
Use Lemma \ref{lemma},
\end{itemize}
to get the decomposition
\[
\begin{split}
&\left(
\begin{array}{cccccc}
\alpha  &\cdot&\cdot    &    \cdot & -(\alpha +tp) e^{i\theta}& *  \\
\cdot&p &v    &    \cdot & \cdot & \cdot    \\
\cdot&\bar v&t^2r    &    \cdot & * & *    \\
\cdot&\cdot&\cdot    &  t^2p & \cdot & *  \\
-(\alpha +tp) e^{-i\theta}&\cdot&*    &    \cdot & \alpha  & \cdot     \\
*& \cdot& *    &    * & \cdot & q
\end{array}
\right)\\
=&
\left(
\begin{array}{cccccc}
\alpha  &\cdot&\cdot    &    \cdot & -\alpha e^{i\theta}& *    \\
\cdot&\cdot &\cdot    &    \cdot & \cdot & \cdot    \\
\cdot&\cdot&*    & \cdot & \cdot & *    \\
\cdot&\cdot&\cdot    &  \cdot & \cdot & \cdot    \\
-\alpha e^{-i\theta}&\cdot&\cdot   &    \cdot & \alpha  & *     \\
*& \cdot   &    *& \cdot & * &*
\end{array}
\right)
+
\left(
\begin{array}{cccccc}
\cdot &\cdot&\cdot    &    \cdot & -tp \, e^{i\theta}& -t v\, e^{i\theta}   \\
\cdot&p &v    &    \cdot & \cdot & \cdot    \\
\cdot&\bar v&*    &   \cdot & * & *    \\
\cdot&\cdot&\cdot    &  t^2p & \cdot & *    \\
-tp\, e^{-i\theta}&\cdot&*   &    \cdot & \cdot  & \cdot     \\
-t\bar v\, e^{-i\theta}& \cdot& *    &    *& \cdot & *
\end{array}
\right).
\end{split}
\]

Now, we use the exactly same argument to the $2\times 2$ principal sub-block
\[
\left(\begin{array}{cc} \phi_{22} & \phi_{23}\\ \phi_{32} & \phi_{33}\end{array}\right)
=\left(\begin{array}{cc}P_{22} & P_{23} \\ P_{32} & P_{33} \end{array}\right)
+\left(\begin{array}{cc} Q_{22} & Q_{23} \\ Q_{32} & Q_{33} \end{array}\right),
\]
to see that
the $(3,3)$ entry of the positive part $P_{22}$ should be zero. From the third and the first blocks
\[
\left(\begin{array}{cc} \phi_{11} & \phi_{13}\\ \phi_{31} & \phi_{33}\end{array}\right)
=\left(\begin{array}{cc} P_{11} & P_{13} \\ P_{31} & P_{33} \end{array}\right)
+\left(\begin{array}{cc} Q_{11} & Q_{13} \\ Q_{31} & Q_{33} \end{array}\right),
\]
we also see that
the $(3,3)$ entry of $P_{11}$ is also zero. Therefore, all entries of the positive part must be zero except
for entries containing $\alpha$. Consequently, we see that
\[
P=\left(\begin{array}{ccc} P_{11} & P_{12} & P_{13}\\P_{21} & P_{22} & P_{23}\\P_{31} & P_{32} & P_{33}\end{array}\right)=
\left(\begin{array}{ccccccccc}
\alpha & \cdot & \cdot &\cdot & -\alpha e^{i\theta}& \cdot & \cdot & \cdot & -\alpha e^{-i\theta}\\
\cdot & \cdot & \cdot & \cdot & \cdot & \cdot & \cdot & \cdot & \cdot \\
\cdot & \cdot & \cdot & \cdot & \cdot & \cdot & \cdot & \cdot & \cdot \\
\cdot & \cdot & \cdot & \cdot & \cdot & \cdot & \cdot & \cdot & \cdot \\
-\alpha e^{-i\theta}& \cdot & \cdot &\cdot & \alpha & \cdot & \cdot & \cdot & -\alpha e^{i\theta}\\
\cdot & \cdot & \cdot & \cdot & \cdot & \cdot & \cdot & \cdot & \cdot \\
\cdot & \cdot & \cdot & \cdot & \cdot & \cdot & \cdot & \cdot & \cdot \\
\cdot & \cdot & \cdot & \cdot & \cdot & \cdot & \cdot & \cdot & \cdot \\
-\alpha e^{i\theta}& \cdot & \cdot &\cdot & -\alpha e^{-i\theta}& \cdot & \cdot & \cdot & \alpha
\end{array}\right).
\]

We proceed to determine the matrix $Q$ with $C_\phi=P+Q$ as follows:
\begin{itemize}
\item
Take the $2\times 2$ sub-blocks from the first and the second blocks, and compare the both side of $C_\phi=P+Q$,
\item
Take the partial transposes of sub-blocks of $Q$,
\item Do the same thing for other $2\times 2$ sub-blocks, to get the partial transpose $Q^\Gamma$ of $Q$,
\item
Apply Lemma \ref{lemma},
\end{itemize}
to get the following:
\[
Q^\Gamma=\left(
\begin{array}{ccccccccc}
\cdot &\cdot&\cdot    &    \cdot & \cdot & \cdot   &\cdot&\cdot&\cdot \\
\cdot&p &v    &    -tp\, e^{-i\theta} & \cdot & - \bar u\,  e^{-i\theta}/t &-v\, e^{i\theta}/t& \bar C &\cdot  \\
\cdot&\bar v&t^2r    &   -t\bar v\, e^{-i\theta}& \cdot & A  &-tr\, e^{i\theta}&-tw\, e^{i\theta}&\cdot\\
\cdot&-tp\, e^{i\theta}&-t v\, e^{i\theta}&  t^2p & \cdot &  \bar u   &B &-t\bar u\, e^{-i\theta}&\cdot\\
\cdot &\cdot&\cdot   &    \cdot & \cdot  & \cdot  &\cdot&\cdot&\cdot   \\
\cdot& - u \,e^{i\theta}/t& \bar A    &     u& \cdot & q &-\bar w \,e^{-i\theta}/t&-tq \, e^{-i\theta}&\cdot\\
\cdot &-\bar v \,e^{-i\theta}/t&-tr \, e^{-i\theta}&\bar B &\cdot &- w\, e^{i\theta}/t&r &w &\cdot\\
\cdot &C &-t\bar w\, e^{-i\theta}&-t u\, e^{i\theta}&\cdot &-tq\, e^{i\theta}&\bar w &t^2q &\cdot\\
\cdot &\cdot&\cdot    &    \cdot & \cdot & \cdot   &\cdot&\cdot&\cdot \\
\end{array}
\right).
\]
Therefore, we have determined $C_\phi$ with
$$
C_\phi=P+Q,
$$
with unknowns $u,v,w,A,B,C$ and $\alpha$.

Now, we consider the following four product vectors
\begin{equation}\label{ten-prod}
\begin{split}
&(1,1,1)^{\rm t} \otimes (1,1,1)^{\rm t},\quad \quad \quad
(1,1,-1)^{\rm t} \otimes (1,1,-1)^{\rm t},\\
&(1,-1,1)^{\rm t} \otimes (1,-1,1)^{\rm t},\ \quad
(-1,1,1)^{\rm t} \otimes (-1,1,1)^{\rm t},
\end{split}
\end{equation}
which are of the form (\ref{eq:case1}), to get an unnormalized separable state summing up pure product states arising from them.
Considering the pairing of this state and $C_\phi=P+Q$, we get
\begin{equation}\label{eq:pos_cond}
\alpha =\frac{ (p+q+r)|t-e^{i\theta}|^2}{3(2\cos\theta -1)}
=\frac{ (p+q+r)|t-e^{i\theta}|^2}{3(p_{\theta} -1)}\ \ {\rm for }\ -\frac{\pi}3<\theta<\frac{\pi}3.
\end{equation}
So, if we take the pairing with a product vector of the form \eqref{eq:case1}, we see that the pairing
\[
\begin{split}
\bar a_1 a_2(A-s_0 w)&+\bar a_2 a_3(B-s_0 v)+\bar a_3 a_1(C-s_0 u)\\
&+a_1 \bar a_2(\bar A-s_0 \bar w)+a_2\bar  a_3(\bar B-s_0 \bar v)+a_3\bar  a_1(\bar C-s_0 \bar u)
\end{split}
\]
must be zero for any $a_1,a_2,a_3$ with $|a_1|=|a_2|=|a_3|$,
where $s_0=(t^2e^{i\theta}-t+e^{i\theta})/t$. By considering the following six cases of $(a_1,a_2,a_3)$:
\[
(1,1,1),\ (1,1,-1),\ (1,1,i),\ (1,i,-1),\ (1,i,i),\ (1,i,-i),
\]
we see that
\begin{equation}\label{nnnnn}
A=\frac {(t^2 e^{i\theta}-t+e^{i\theta})}t w,\ B=\frac {(t^2 e^{i\theta}-t+e^{i\theta})}t v,
\ C=\frac {(t^2 e^{i\theta}-t+e^{i\theta})}t u.
\end{equation}

In order to determine $u,v$ and $w$, we proceed as follows:
\begin{itemize}
\item
Take $6\times 6$ principal submatrix of $C_\phi=P+Q$ with $1,2,4,5,7,8$ columns and rows, which
represents a decomposable map from $M_3$ to $M_2$,
\item
Look at the $3\times 3$ principal submatrix of the positive matrix part from $1,4,5$ and $1,4,6$ columns and rows,
\item
Take the partial transpose of the co-positive matrix part, and look at the $3\times 3$ principal submatrix from $2,3,6$ columns and rows,
\end{itemize}
to get the inequality
$-t^2p|C-ue^{2 i\theta}|^2\ge 0$.
Therefore, we have $C= u e^{2i\theta}$, and
\[
[(t+1/t)e^{i\theta}-(1+e^{2i\theta})]u=0
\]
by (\ref{nnnnn}). Since $|1+e^{2i\theta}|<|(t+1/t)e^{i\theta}|$ for $(\theta,t)\neq (0,1)$, we get $u=0$.
By the exactly same way, we also see that $v=w=0$.
Therefore, the matrices representing the positive maps in
the double dual face of $\Phi_\theta(t)$ are of the form
\[
\left(
\begin{array}{ccccccccc}
\alpha  &\cdot&\cdot    &    \cdot & -(\alpha +tp) e^{i\theta}& \cdot    &    \cdot &\cdot & -(\alpha +tr) e^{-i\theta}\\
\cdot&p &\cdot    &    \cdot & \cdot & \cdot    &    \cdot &\cdot & \cdot\\
\cdot&\cdot&t^2r    &    \cdot & \cdot & \cdot    &    \cdot &\cdot & \cdot\\
\cdot&\cdot&\cdot    &  t^2p & \cdot & \cdot    &    \cdot &\cdot & \cdot\\
-(\alpha +tp) e^{-i\theta} &\cdot&\cdot    &    \cdot & \alpha  & \cdot    &    \cdot &\cdot & -(\alpha +tq) e^{i\theta}\\
\cdot&\cdot&\cdot    &    \cdot & \cdot & q    &    \cdot &\cdot & \cdot\\
\cdot&\cdot&\cdot    &    \cdot & \cdot & \cdot    &    r &\cdot & \cdot\\
\cdot&\cdot&\cdot    &    \cdot & \cdot & \cdot    &    \cdot &t^2q & \cdot\\
-(\alpha +tr) e^{i\theta}&\cdot&\cdot    &    \cdot & -(\alpha +tq) e^{-i\theta} & \cdot    &    \cdot &\cdot & \alpha  \end{array}
\right),
\]
for nonnegative real numbers $p,\,q,\,r$.

Note that the corresponding map sends $[x_{ij}]$ to
\[
\left(\begin{array}{ccc}
\alpha x_{11}+t^2px_{22}+rx_{33} & -(\alpha +tp) e^{i\theta}x_{12}  & -(\alpha +tr) e^{-i\theta}x_{13}\\
-(\alpha +tp) e^{-i\theta}x_{21} & \alpha x_{22}+t^2qx_{33}+px_{11}  & -(\alpha +tq) e^{i\theta}x_{23}\\
-(\alpha +tr) e^{i\theta}x_{31} & -(\alpha+tq) e^{-i\theta}x_{32} &  \alpha x_{33}+t^2rx_{11}+qx_{22}
\end{array}\right).
\]
By the relation (\ref{eq:Choi_pair}) and product vectors of the form (\ref{eq:case1}), we see that
the above matrix is singular with $x_{ij}=1$ for $i,j=1,2,3$. By a direct calculation, the determinant
is given by
\[
\begin{split}
D[\alpha,p,q,r,\theta]=&S_3t^6+\alpha S_2t^4-2(\alpha  S_2+\cos (3\theta) S_3)t^3
+\alpha (1-2\cos(3\theta))S_2 t^2 \\
&\hskip 1pt
 -2\alpha (\alpha S_1+\alpha\cos(3\theta)S_1+S_2)t +S_3 +\alpha S_2 -2\alpha^3(1+\cos(3\theta))= 0,
\end{split}
\]
with
\[
S_1=p+q+r, \qquad S_2=pq+qr+rp, \qquad S_3=pqr.
\]

For simplicity, we denote $T_1$, $T_2$ and $T_3$ by
\[
T_1:=-\frac 4{27}S_1^3+\frac 13 S_1 S_2+S_3, \quad
T_2:=-\frac 49 S_1^3 + \frac 43 S_1 S_2,\quad T_3:=\frac 49 pqr -\frac{4}{27}(p^3+q^3+r^3).
\]
Since  $\alpha=  |t-e^{i\theta}|^2 S_1/(6\cos\theta-3)$ from \eqref{eq:pos_cond}, we can write
\[
0=D[\alpha,p,q,r,\theta]=f_1(t,\theta)T_1 -\frac {f_1(t,\theta)-f_2(t,\theta)}4 T_2
\]
with
\[
\begin{split}
f_1(t,\theta)=&(1-2t^3\cos (3\theta)+t^6)=|t^3-e^{3i\theta}|^2,\\
f_2(t,\theta)=&\frac{|t-e^{i\theta}|^2(1-2t+t^2-2t^3+t^4-2t^2\cos (3\theta))}{2\cos\theta-1}.
\end{split}
\]
We note that $f_1(t,\theta)>0$ by \eqref{cond}, with which one can also show the following inequality:
\begin{equation}\label{f_cond}
f_1(t,\theta)+3 f_2(t,\theta)=\frac{4 \cos^2 (\theta/2) |t-e^{i\theta}|^4\left(t^2+(4\cos\theta -3)t+1\right)}{2\cos\theta-1}> 0.
\end{equation}
On the other hand, we also have
\[
\begin{split}
& T_1\le T_3\le 0,\\
&T_1-T_2\le  -2T_3\\
&T_1-\frac{T_2}4 =-\frac 1{27}(p+q+r)^3+pqr\le 0,
\end{split}
\]
where the equalities hold if and only if $p=q=r$ in any cases.
We also have
\[
T_2=-\frac{2}{9} (p+q+r)\left( (p-q)^2+(q-r)^2+(r-p)^2\right)\le 0.
\]

Now, we proceed to show that $D[\alpha,p,q,r,\theta]=0$ implies that $p=q=r$.
In the case of $f_2(t,\theta)\ge 0$,  we see that $p=q=r$  by the following inequality:
\[
0\le D[\alpha,p,q,r,\theta]=f_1(t,\theta)\left (T_1-\frac {T_2}4\right)+\frac{f_2(t,\theta)}4 T_2\le 0.
\]
In the case of $f_2(t,\theta)<0$, we see that $f_1(t,\theta)-f_2(t,\theta)> 0$ and
\[
3f_1(t,\theta)+f_2(t,\theta)=3[f_1(t,\theta)+3f_2(t,\theta)]-8f_2(t,\theta)>0
\]
by \eqref{f_cond}. Therefore, we get $p=q=r$ by the following inequality:
\[
\begin{split}
0\le D[\alpha,p,q,r,\theta]&=\frac{f_1(t,\theta)-f_2(t,\theta)}4 (T_1-T_2)+\frac {3f_1(t,\theta)+f_2(t,\theta)}4 T_1\\
&\le  \frac{f_1(t,\theta)-f_2(t,\theta)}4 (-2T_3)+\frac {3f_1(t,\theta)+f_2(t,\theta)}4 T_3\\
&= \frac{f_1(t,\theta)+3f_2(t,\theta)}4 T_3\le 0.
\end{split}
\]
Consequently, we can conclude that $p=q=r$ and  $\alpha=|t-e^{i\theta}|^2 p/(2\cos\theta-1)$. We note that
\[
\alpha+tp=\left(\frac{|t-e^{i\theta}|^2+(2\cos\theta-1)t}{2\cos\theta-1} \right)p=\frac{(t^2-t+1)p}{2\cos\theta-1}.
\]
Then, the corresponding positive maps are the scalar multiples of the Choi type map
\[
\frac{(1-t+t^2)p}{2\cos\theta-1}\,\Phi[a(t),b(t),c(t);\theta]
\]
since $p_{\theta}=2\cos\theta$ for $|\theta|<\pi/3$.
This completes the proof that the positive map $\Phi_\theta(t)$ generates an exposed ray under the condition \eqref{cond}.

Now, we show that the map $\Phi_\theta(t)$ is exposed under the condition (\ref{cond-exp}), in general. To do this, we consider the diagonal
unitary matrix $U={{\rm Diag}}(1,e^{-\frac 23\pi i},e^{-\frac 43\pi i})$, and
define the affine isomorphism $\Lambda$ on the convex cone $\mathbb P$ by
\[
(\Lambda \phi)(X)=U^*\phi(X)U,\qquad X\in M_3.
\]
Then we have
\[
\Lambda[\Phi_\theta(t)]=\Phi_{\theta-\frac 23\pi}(t),
\]
since $p_\theta=p_{\theta\pm \frac 23\pi}$. Because exposedness is invariant under the affine
isomorphism $\phi\mapsto \Lambda\phi$, we conclude the following:

\begin{theorem}
The positive map $\Phi_\theta(t)$ is exposed under the condition {\rm (\ref{cond-exp})}.
\end{theorem}

\section{Lengths of separable states}

In this section, we find a $3\otimes 3$ separable state with length ten in the dual face of the map $\Phi_\theta(t)$.
We use the product vectors  $|z_1(a_1,a_2,a_3)\rangle$ and $|z_i(\beta)\rangle$ of the form
\eqref{eq:case1}, \eqref{eq:case2}, \eqref{eq:case3} and \eqref{eq:case4},
to construct the following unnormalized separable state
\[
\begin{split}
&\varrho_0(\theta,t)
=\frac 1{16}\sum_{k,\ell=1}^4 |z_1(1,i^k,i^{\ell})\rangle \langle z_1(1,i^k,i^\ell)|
+\frac 1{3t^2}\sum_{k=2}^4\sum_{\ell=1}^3 |z_k(e^{2i\pi\ell/3})\rangle \langle z_k(e^{2i\pi\ell/3})|\\
=&\left(\begin{array}{ccccccccc}
3 & \cdot & \cdot & \cdot & 1+e^{-i \theta} & \cdot & \cdot & \cdot & 1+e^{i\theta}\\
\cdot & 1+t & \cdot & \cdot & \cdot & \cdot & \cdot & \cdot & \cdot \\
\cdot & \cdot & 1+1/t & \cdot & \cdot & \cdot & \cdot & \cdot & \cdot \\
\cdot & \cdot & \cdot & 1+1/t & \cdot & \cdot & \cdot & \cdot & \cdot \\
1+e^{i\theta} & \cdot & \cdot & \cdot & 3 & \cdot & \cdot & \cdot & 1+e^{-i\theta}\\
\cdot & \cdot & \cdot & \cdot & \cdot & 1+t & \cdot & \cdot & \cdot \\
\cdot & \cdot & \cdot & \cdot & \cdot & \cdot & 1+t & \cdot & \cdot \\
\cdot & \cdot & \cdot & \cdot & \cdot & \cdot & \cdot & 1+1/t & \cdot \\
1+e^{-i\theta} & \cdot & \cdot & \cdot & 1+e^{i \theta} & \cdot & \cdot & \cdot & 3\\
\end{array}\right).
\end{split}
\]

It is clear that $\varrho_0:=\varrho_0(\theta,t)$ is in the dual face of the map $\Phi_\theta(t)$. Actually,
$\varrho_0$ is an interior point of the dual face $\Phi_\theta(t)^\prime$.
To see this, we first note that every extreme ray of $\Phi_\theta(t)'$ is
generated by a pure product state $|z\rangle \langle z|$ determined by one of the product vectors
\[
|z_1(1,\beta,\gamma)\rangle,\qquad |z_2(\beta)\rangle,\qquad |z_3(\beta)\rangle,\qquad |z_4(\beta)\rangle
\]
with $|\beta|=|\gamma|=1$. We also note that
\[
\varrho_0=\frac 1{16}\sum_{k,\ell=1}^4 |z_1(1,i^k\beta,i^{\ell}\gamma\rangle \langle z_1(1,i^k\beta,i^\ell\gamma)|
+\frac 1{3t^2}\sum_{k=2}^4\sum_{\ell=1}^3 |z_k(\beta e^{2\pi i\ell/3})\rangle \langle z_k(\beta e^{2\pi i\ell/3})|,
 \]
by a direct computation. Therefore, we see that for any $\rho$ in $\Phi_\theta(t)'$ generating an extreme ray,
there exist a state $\sigma$ in $\Phi_\theta(t)'$ such that $\varrho_0=\rho+\sigma$. This tells us that $\rho_0$ is an interior point
of the face $\Phi_\theta(t)^\prime$.

In order to determine the length of $\varrho_0$, we note that any possible decomposition of $\varrho_0$ into the sum of
pure product states is given by
\begin{equation}\label{eq:sum}
\varrho_0=\sum_{k=1}^4 \sum_{i=1}^{n_k}t_{ki} W_{ki}
\end{equation}
with
\[
\begin{split}
W_{1i}&= |z_1(1,\beta_{1i},\gamma_{1i})\rangle \langle z_1(1,\beta_{1i},\gamma_{1i})|,\\
W_{ki}&=\frac 1{t^2}|z_{k}(\beta_{ki})\rangle \langle z_{k}(\beta_{ki})|, \qquad k=2,3,4,
\end{split}
\]
for complex numbers $\beta_{ki}$ and $\gamma_{1i}$ with modulus one.
Therefore, the length $\ell(\varrho_0)$ is given by
\[
\ell(\varrho_0)=\min\{n_1+n_2+n_3+n_4\,:\, \rho_0=\sum_{k=1}^4 \sum_{i=1}^{n_k}t_{ki} W_{ki},\ t_{ki}>0\}.
\]
Comparing the diagonal entries of \eqref{eq:sum}, we see that $\sum_{i=1}^{n_k} t_{ki}=1$, and so  $n_k\ge 1$ for $k=1,2,3,4$.
We also compare the $(1,2)$, $(2,5)$ and $(4,5)$ entries in the identity \eqref{eq:sum}, to get
$\sum_{i=1}^{n_1} t_{1i} \beta_{1i}=\sum_{i=1}^{n_4} t_{4i} \beta_{4i}=0$. Comparing various entries, we
get the relations
\begin{equation}\label{rel}
\sum_{i=1}^{n_1} t_{1i} \beta_{1i}
=\sum_{i=1}^{n_1}  t_{1i} \gamma_{1i}
=\sum_{i=1}^{n_2} t_{2i} \beta_{2i}
=\sum_{i=1}^{n_3} t_{3i} \beta_{3i}
=\sum_{i=1}^{n_4} t_{4i} \beta_{4i}=0.
\end{equation}
From this, we see that $n_k\ge 2$ for $k=1,2,3,4$. If $n_1=2$ then we have
\[
t_{11}=t_{12}=\frac 12,\qquad \beta_{11}+\beta_{12}=\gamma_{11}+\gamma_{12}=0,
\]
which is not possible by the $(1,6)$ entries of \eqref{eq:sum}. Therefore, we conclude that
$n_1\ge 3$.

In order to show that $\ell(\varrho_0)>9$, we assume that $\ell(\varrho_0)=9$ then we have
$n_1=3$ and $n_2=n_3=n_4=2$, and so we also have the relations
\[
\beta_{21}+\beta_{22}=\beta_{31}+\beta_{32}=\beta_{41}+\beta_{42}=0.
\]
By comparing the $(2,4)$-entries of the matrices in \eqref{eq:sum}, we have the identity
\[
\beta_{41}^2e^{i\theta}=-\sum_{i=1}^{3}t_{1i}\bar \beta_{1i}^2.
\]
So we get the inequality
\[
1=|\beta_{41}^2|
= |\sum_{i=1}^{3} t_{1i} \bar \beta_{1i}^2|\le \sum_{i=1}^{3} t_{1i}|\bar \beta_{1i}^2|=\sum_{i=1}^{3}t_{1i}=1.
\]
Then we conclude that  $\beta_{11}^2=\beta_{12}^2=\beta_{13}^2$
from the equality condition of the triangle inequality.
By re-indexing, we may write $\beta_{1i_1}=\beta_{1i_2}=-\beta_{1i_3}$ and have the relation
\[
\bar \beta_{1i_1}\left (t_{1i_1}\gamma_{1i_1}+t_{1i_2}\gamma_{1i_2}-t_{1i_3}\gamma_{1i_3}\right)=0,
\]
from the $(2,3)$ entries. Therefore, we see that $\gamma_{1i_3}=0$ from the relations \eqref{rel}.
This contradicts to the condition $|\gamma_{1i_3}|=1$, and
completes the proof of $\ell(\rho)>9$.

Finally, we exhibit an explicit decomposition for $\varrho_0$ with $n_1=4$ and $n_2=n_3=n_4=2$
to show that $\ell(\varrho_0)=10$. To do this, we take $W_{1j}\, (j=1,2,3,4)$ determined by the following $(\beta_{1j},\gamma_{1j})$:
\[
\begin{split}
(\beta_{11},\gamma_{11})=&(e^{i \theta} i, e^{i\theta/2} i),\quad \quad  (\beta_{12},\gamma_{12})=(-e^{i \theta} i, e^{i\theta/2} i),\\
(\beta_{13},\gamma_{13})=&(e^{i \theta} i, -e^{i\theta/2} i),\ \quad  (\beta_{12},\gamma_{12})=(-e^{i \theta} i, -e^{i\theta/2} i).
\end{split}
\]
We also take $W_{k1}$ and $W_{k2}\, (k=2,3,4)$ determined by $\beta_{k1}$ and $\beta_{k2}$, respectively:
\[
(\beta_{21},\beta_{22})=(i,-i),\quad (\beta_{31},\beta_{32})=(1,-1),\quad
(\beta_{41},\beta_{42})=(e^{-3i \theta/2},-e^{-3i \theta/2}).
\]
Then, by a direct computation, we see that
\[
\varrho_0 = \frac 14 (W_{11}+W_{12}+W_{13}+W_{14})+\frac 1 2(W_{21}+W_{22}+W_{31}+W_{32}+W_{41}+W_{42}),
\]
and this complete the proof of $\ell(\varrho_0)=10$.

Now, we consider the following unnormalized separable state
$$
\varrho_1(\theta,t)=\varrho_0(\pi-\theta,t)+2W[2,t,1/t;0].
$$
This is separable, because it was shown in \cite{kye_osaka} that $W[a,b,c;0]$ is separable if and only if it is of PPT.
By a direct computation, we see that $\varrho_0(\theta,t)+\varrho_1(\theta,t)$ is a diagonal matrix.
This shows that
$\ell(\varrho_0)=10$ but $\ell(\varrho_0+\varrho_1)=9$.

In \cite{ha+kye_unique_decom}, the authors exhibited an example of a separable state
of length $10$ which has a unique decomposition. Our construction
also gives rise to such examples. To see this, we first note the following relation
\[
\xi\otimes\eta\in P_\phi\ \Longleftrightarrow\ \bar\xi\otimes\eta\in P_{\phi\circ {\rm t}}
\]
for a positive map $\phi$. Furthermore, we have
$P_{\phi+\psi}=P_\phi\cap P_\psi$. Therefore, we see that the set $P_{\phi+\phi\circ {\rm t}}$ consists of exactly ten
product vectors
\[
z_1(1,1,1),\ z_1(-1,1,1),\ z_1(1,-1,1),\ z_1(1,1,-1),\
z_2(\pm 1),\ z_3(\pm 1),\  z_4(\pm 1),
\]
for $\phi=\Phi_\theta(t)$.
Note that the first four product vectors appear in (\ref{ten-prod}). It is easy to see that these ten product vectors
give rise to linearly independent pure product states. Therefore, these ten pure product states generate a simplicial face
of the convex set of all separable states by \cite{ha+kye_unique_decom}, which is affinely isomorphic to the $9$-dimensional simplex
with ten vertices. Every separable state in this face has a unique decomposition.


The authors are grateful to Lin Chen, Xin Li, \L ukasz Skowronek and Karol Zyczkowski for discussions on the topics.
This work was partially supported by the Basic Science Research Program through the
National Research Foundation of Korea(NRF) funded by the Ministry of Education, Science
and Technology (Grant No. NRFK 2012-0002600 to K.-C. Ha and Grant No. NRFK 2012-0000939 to S.-H. Kye)








\end{document}